\documentclass[twocolumn,showpacs,prl]{revtex4}

\usepackage{epsfig}

\begin{document}

\title{Fast crack propagation by surface diffusion}

\author{Efim A. Brener}
\author{Robert Spatschek}
\affiliation{Institut f\"ur Festk\"orperforschung, Forschungszentrum J\"ulich, D-52425 J\"ulich, Germany}

\date{\today}

\begin{abstract}
We present a continuum theory which describes the fast growth of a crack by surface diffusion.
This mechanism overcomes the usual cusp singularity by a self-consistent selection of the crack tip radius.
It predicts the saturation of the steady state crack velocity appreciably below the Rayleigh speed and tip blunting.
Furthermore, it includes the possibility of a tip splitting instability for high applied tensions.
\end{abstract}


\maketitle

One of the most challenging puzzles in nonequilibrium physics and materials science is the phenomenon of fracture.
It is important for the vast field of material failure and probably also for friction processes \cite{Gerde01}.
Despite its relevance even the motion of a single crack is poorly understood \cite{Hellemans98}. 
Experimentally, the maximum attained crack velocities are far lower than the theoretically expected Rayleigh speed \cite{Freund90}.
Beyond a critical velocity, a so far unpredictable tip splitting of the crack can happen and produce strange oscillations of the crack speed \cite{Sharon99}.

\begin{figure}
\epsfig{file=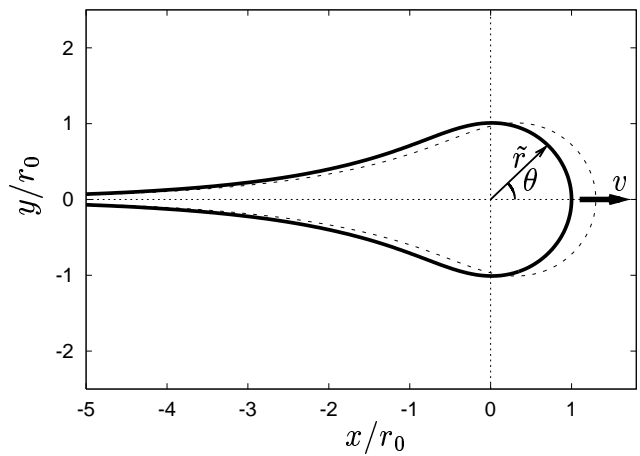, width=8.5cm}
\caption{Calculated shape of the crack (without elastic displacements) driven by surface diffusion for $\Delta=2$. The advance of the crack in positive $x$ direction is indicated by the dashed curve. This requires the redistribution of matter along the crack by a transport mechanism.}
\label{Fig1}
\end{figure}

The classical theories \cite{Freund90} are based on linear theory of elasticity and on an integral energy balance in the vicinity of an infinitely sharp crack tip.
However, a more detailed approach based on equations of motion for the crack shape is needed to describe the intriguing spectrum of phenomena near the crack tip.
Hence, the curvature of the crack tip is required as a new relevant dynamical variable which also allows to avoid stress singularities.
We emphasize that, in contrast to models which describe crack propagation by bond breaking at the infinitely sharp tip, growth with a finite tip radius always requires a transport mechanism in order to preserve the shape (see Fig. \ref{Fig1}).
Recently it was proposed that this lengthscale is dynamically selected by the threshold of plastic deformations in the tip region \cite{Langer00}.
Unfortunately, approaches of this type (see also \cite{Aranson00}) require the introduction of theories of plasticity which are usually much more speculative and less verified than the ordinary linear theory of elasticity.

Here we demonstrate that {\em linear theory of elasticity} is sufficient to describe consistently crack propagation, driven by {\em surface diffusion} along the crack surfaces.
Of course, in many situations plasticity is very important, but the beauty of our approach is that it predicts, in a simple and well controlled continuum theory, steady state crack growth, the tip splitting instability and also slow deformations of already existing cracks.

The idea of crack propagation by surface diffusion has previously been studied by Stevens and Dutton \cite{Stevens71}, who assumed an ad hoc crack shape which is not found by solving the full free boundary problem and requires mass transport over large scales.
Therefore, their model cannot describe the usual fast crack growth.

Our basic mechanism  is related to the Asaro-Tiller-Grinfeld (ATG) instability \cite{Asaro72}, which predicts a morphological instability of a uniaxially stressed solid interface due to surface diffusion.
Relatively long-wave perturbations of the interface lead to a reduction of the elastic energy of the system, whereas short-wave corrugations are hampered by surface energy.
In the long time behavior, deep grooves can form, producing shapes similar to cracks \cite{Yang93,Spencer94}.
According to previous theories, which used only the {\em static} theory of elasticity, the notches propagate with increasing velocity and decreasing tip radius and collapse to a finite-time cusp singularity.
Similar to crack dynamics the lack of tip radius selection becomes obvious, and already shows the close relationship between the ATG instability and crack propagation.

Usually, it is believed that surface diffusion is slow, but, surprisingly enough, it should not be ignored even in fast fracture processes.
Our main idea is that surface diffusion is driven by the strong gradient of the chemical potential in the tip region.
This can be a very efficient mechanism for crack propagation if the transport length is sufficiently small.
Additionally, energy release and strong dissipation bring the local temperature close to the melting temperature \cite{Fuller75,Fineberg99}.
This drastically increases the surface diffusion coefficient and makes fast crack propagation essentially independent of the outside temperature.

{\em Steady state crack growth ---}
We use a two-dimensional plane-strain situation with mode I loading to describe crack propagation \cite{Freund90}.

On the surfaces of the crack the normal stress $\sigma_{nn}$ and the shear stress $\sigma_{n\tau}$ vanish, whereas the tangential stress $\sigma_{\tau\tau}$ usually does not.
The chemical potential at the interface is given by
\begin{equation} \label{Eq1}
\mu = \Omega \left(\frac{1-\nu^{2}}{2E} \sigma_{\tau\tau}^{2} - \alpha \kappa\right) .
\end{equation}
Here $\alpha$ is the surface energy, $\kappa$ the curvature of the interface and $\Omega$ the atomic volume.
$E$ and $\nu$ are Young's modulus and Poisson ratio, respectively.
Non-hydrostatic stresses drive a surface flux proportional to the gradients of the chemical potential along the surface;
in turn the normal velocity equals the divergence of this flux due to conservation of material, 
\begin{equation} \label{Eq2}
v_{n} = -\frac{D}{\alpha \Omega} \frac{\partial^{2}\mu}{\partial s^{2}}
\end{equation}
where $\partial/\partial s$ denotes the tangential derivative and $D$ (dimension $\rm m^{4}s^{-1}$) is proportional to the surface diffusion coefficient. (It is related to the usual surface diffusion coefficient $D_{s}$ by $D=D_{s}\Omega^{2}\delta\, \alpha/kT$. Here $\delta$ is the number of atoms per unit area of surface, $k$ the Boltzmann constant and $T$ the temperature.)

First, we are interested in steady state solutions of the equation of motion, with a crack moving in positive $x$-direction with  velocity $v$ (see Fig. 1);
In co-moving polar coordinates, $x=r(\theta)\cos\theta, y=r(\theta)\sin\theta$, the steady state equation for the shape $r(\theta)$ reads after one integration of Eq. (\ref{Eq2})
\begin{equation} \label{Eq3}
vr\sin\theta = -\frac{D}{\alpha \Omega}\frac{1}{\sqrt{r^{2}+r'^{2}}} \frac{d\mu}{d\theta}.
\end{equation}
Generally speaking, this, together with Eq. (\ref{Eq1}), is a complicated, non-linear third order equation with non-local contributions arising from the elastic fields, since $\sigma_{\tau\tau}$ depends on the entire shape.

In the tail region stresses decay and the shape equation is a third order linear differential equation $D y''' = v y$ with two growing (and oscillating) and one decaying solution.
Only the latter, $y(x\to -\infty)=A\exp[(v/D)^{1/3}x]$, asymptotically describes physical shapes and is allowed.
Let us focus on symmetrical solutions, $r(\theta)=r(-\theta)$, and start integration at the crack tip $\theta=0$.
Since the physical properties, curvature and stresses, do not depend on the choice of coordinate system but only on the crack shape, we can arbitrarily chose $r(\theta=0)=r_{0}$, with the a priori unknown tip radius $r_{0}=1/\kappa(0)$.
Then from symmetry and the definition of the tip curvature, $\kappa=(r^{2}+2r'^{2}-rr'')/(r^{2}+r'^{2})^{3/2}$, the natural conditions $r'(0)=r''(0)=0$ arise.
Integration over the upper interface $\theta>0$ requires the suppression of two growing exponentials at the tail, which imposes two boundary conditions.
For a given external loading, these conditions can be fulfilled by a proper selection of the tip radius $r_{0}$ and growth velocity $v$.
By this argument the situation seems to be fully described.
However, as we mentioned already earlier, the use of a {\em static} theory of elasticity does not allow a selection of the tip radius.
The reason is, that both contributions to the chemical potential, surface energy $\mu_{s} \sim \kappa$ and elastic energy $\mu_{el}\sim \sigma^{2}$, behave as $r_{0}^{-1}$ close to the tip:
In the tip approximation, stresses behave as \cite{Freund90}

\begin{equation} \label{Eq4}
\sigma_{ij}= \frac{K}{r^{1/2}}f_{ij}(\theta)
\end{equation}
with the static stress intensity factor $K\sim \sigma_{\infty} L^{1/2}$, where $\sigma_{\infty}$ is the applied remote stress and $L$ is the macroscopic length of the crack, $L\gg r_{0}$, which is not considered here.
Instead we assume $K$ to be kept fixed.
The universal stress distribution $f_{ij}$ depends only on the orientation relative to the crack \cite{Freund90}.
The asymptotic distribution (\ref{Eq4}) is valid far away from the tip, $r_{0}\ll r \ll L$, but it gives the correct scaling of stresses also on the crack surface $r\approx r_{0}$.
Therefore, a dimensionless rescaling of all lengthscales, e.g.  $\tilde{r} = r/r_{0}$, and of the growth velocity $\tilde{v}=v r_{0}^{3}/D$ leaves the equation of motion invariant and, thus, cannot determine the lengthscale $r_{0}$.
Consequently, a steady state solution does not exist.
This is the reason for the already mentioned cusp singularity of the ATG instability.

The main idea of this paper is based on the fact that a full {\em elastodynamic} description restores the selection of this lengthscale. 
It is known that at least for higher crack speeds the angular distribution $f_{ij}$ become strongly dependent on the ratio $v/v_{R}$ ($v_{R}$ is the Rayleigh speed \cite{Freund90}).
The dynamical stress intensity factor $K_{dyn}$ is related to the static one used here by an extra velocity dependent function $g(v/v_{R})$,  $K_{dyn} = K g(v/v_{R})$.
The crucial observation is, that velocity appears now in two different combinations in the equation of motion, $v r_{0}^{3}/D$ and $v/v_{R}$.
Thus, by introduction of the new parameter $v/v_{R}$, a selection of both $v$ and $r_{0}$ happens.

From these general arguments we conclude that fast steady state crack propagation by surface diffusion is indeed possible.
However, the exact solution of the problem is technically very difficult, because it requires the solution of an elastodynamic problem for an a priori unknown crack shape.
The bulk equations of elasticity
\begin{equation} \label{eq5a}
\frac{\partial \sigma_{ij}}{\partial x_{j}}=\rho \ddot{u}_{i}
\end{equation} 
are subject to the boundary conditions on the crack surface (surface of discontinuity) \cite{Freund90} 
\begin{equation} \label{eq6a}
\sigma_{in}+\rho\dot{u}_{i} v_n=0
\end{equation}
with the mass density $\rho$ and the elastic displacement field $u_{i}$.
This is just the momentum balance equation on the free surface which moves with normal velocity $v_n$.
Finally, the expression for the chemical potential on the crack surface should also be corrected compared to the static case: 
\begin{equation} \label{eq7a}
\mu= \Omega \left( \frac{1}{2}\sigma_{ik}u_{ik} - \frac{1}{2}\rho\dot{u}_i^2 - \alpha\kappa \right) 
\end{equation} 
Eqs. (\ref{eq5a})-(\ref{eq7a}) together with surface diffusion Eq. (\ref{Eq2}) and given loading configuration describe the crack propagation in our model.
We note that inertial effects appear not only in the bulk equations of elasticity but also in the boundary conditions and in the expression for the chemical potential.
All these effects lead to the appearance of the parameter $(v/v_R)^2$ in the problem compared to the quasistatic description. 

The preceding equations (\ref {eq5a})-(\ref{eq7a}) can be derived from the Lagrangian
\begin{equation}
{\cal L} = \int_{V(t)} \left( \frac{1}{2}\rho\dot{u}_{i}^{2} - \frac{1}{2}\sigma_{ij} u_{ij} \right) dV - \int_{S(t)} \alpha dS.
\end{equation}
with $V(t)$ being the time dependent volume of the solid and $S(t)$ its surface.
The elastic equations and boundary conditions follow by the condition that ${\cal S}=\int{\cal L}dt$ is stationary with respect to variations of the displacement $u_{i}$, and the chemical potential is related to the variation of ${\cal L}$ with respect to the interface position \cite{Uwaha86, Unpub}.
Evaluation of the exact equations of motion requires extended numerics, especially in the time-dependent case.

{\em The local crack tip model ---}
We simplify the problem in order to make further analytical progress and to expose the general idea of our approach.
It will turn out that one cannot describe all effects by this approximation and further refinement is necessary, but the main results are qualitatively very robust against changes of the model.

We mimic the tangential stress by a {\em local} description in the spirit of Eq. (\ref{Eq4}),
as depending on the propagation velocity and only on the local properties of the interface.
It takes both the velocity dependence of the angular distribution and the decrease of the dynamical stress intensity factor into account:
\begin{equation} \label{Eq5}
\sigma_{\tau\tau}= K \left[ \sqrt{1-(v/v_{R})^{2}}\cos (\theta/2) + (v/v_{R})^{2} \sin^{4}\theta \right] / r^{1/2}.
\end{equation}
This form reflects the first order transition of the principal stress direction $\theta=0$ for low velocities towards $\theta\neq 0$ as function of $v/v_{R}$ \cite{Freund90}.
The use of more sophisticated expressions (e.g. the singular dynamical field in full detail) would not provide a large gain, since, anyway local approximations cannot lead to exact results.
For the same  reasons we also neglect inertial corrections               
to the boundary conditions in (\ref{eq6a}) and in the chemical potential (\ref{eq7a}).

However, we have checked that a model with a continuous transition in azimuthal stress (replacement of $\sin^{4}\theta$ by $\sin^{2}\theta$ in Eq. \ref{Eq5}) gives qualitatively the same results.

\begin{figure}
\epsfig{file=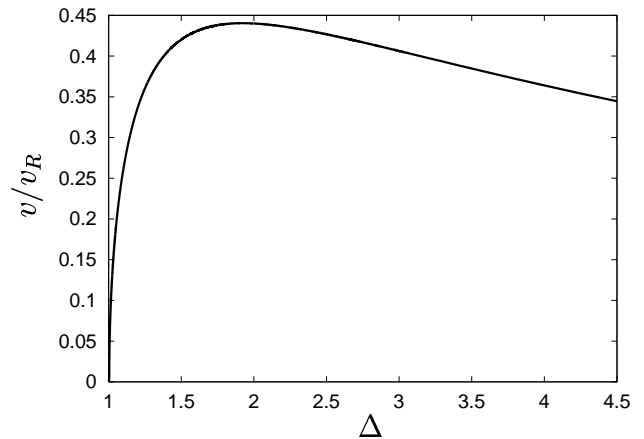, width=8.5cm}
\caption{Steady state velocity of the crack versus dimensionless driving force $\Delta$.}
\label{Fig2}
\end{figure}

\begin{figure}
\epsfig{file=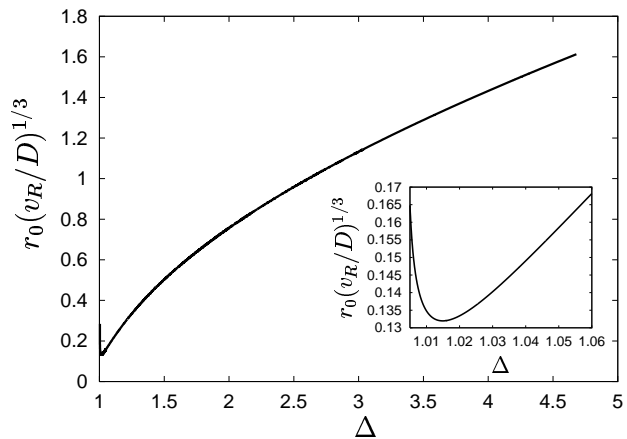, width=8.5cm}
\caption{Dimensionless crack tip radius $r_{0}$ versus dimensionless driving force $\Delta$.}
\label{Fig3}
\end{figure}

Now Eq. (\ref{Eq3}), together with (\ref{Eq1}) and (\ref{Eq5}), is a closed third order differential equation for the shape $r(\theta)$ which can be easily integrated numerically.
It provides both the crack shape (Fig. \ref{Fig1}) and a selection of $v/v_{R}$ and $r_{0}$ as functions of the dimensionless driving force $\Delta=K^{2}(1-\nu^{2})/2E\alpha$.
The results are given in Figures \ref{Fig2} and \ref{Fig3}.

One of the main results is that the upper limit for the steady state crack velocity is appreciable below the Rayleigh speed, as known from experimental results.
The instantaneous velocity in the non steady state regime can of course reach higher values \cite{Sharon99,Fineberg99}.
For relatively low driving forces, the growth velocity increases with increasing $\Delta$,
but for higher values of $\Delta$, it even decreases.
Simultaneously, the tip becomes sharper at first, but then blunts again.

{\em Stability ---}
Although the decrease of the velocity $v$ as function of $\Delta$ might be naively understood as a sign of instability, the model itself is stable:
We performed a straightforward but tedious numerical stability analysis and found no unstable modes.
The point is, that only $v/v_{R}$ decreases, but $vr_{0}^{3}/D$ increases.
Only a decrease of this parameter, which appears in combination with the dissipative coefficient $D$, would be a real sign of instability.

Nevertheless, our solution is subject to the ATG instability above a critical threshold of the driving force $\Delta$.
The de-stabilizing effects stem from the non-local elastic contributions which we neglected in our model, but which are obviously present in the real problem.
This can be proved by simple dimensional analysis arguments:
The characteristic wavelength of the ATG instability is $\lambda\sim E\alpha/\sigma_{\tau\tau}^{2}$ \cite{Asaro72};
in the tip region it reads $\lambda^{(tip)}\sim r_{0}/\Delta$.
Thus, as soon as a certain critical driving force $\Delta_{c}$ is exceeded, the characteristic wavelength of instability fits into the tip region.
The material independent number $\Delta_{c}$ is the threshold for the instability of the steady state solution, whereas $\Delta=1$ is the Griffith point.
Since, according to the steady state solution, $v/v_{R}$ is a universal function of the dimensionless parameter $\Delta$, the threshold of instability in terms of $v/v_{R}$ is also essentially material independent.
It is important to note that our steady state predictions are valid only below the threshold of instability.
Thus, the main part or all of the decrease of velocity versus $\Delta$ is screened by the instability.
Beyond the instability point, the behavior of the system is governed by the full time dependent evolution.
Hence we expect the ATG instability to be the relevant mechanism for the experimentally observed microbranching instability \cite{Sharon99,Fineberg99}.
In contrast to the long wave instability \cite{Brener98}, this instability is localized in the tip region and cannot be suppressed by convective effects.

{\it Conclusion} --- We have developed a self-consistent continuum model for crack propagation in homogeneous media.
Both ingredients of our theory, the {\it linear elasticity} which is valid everywhere in the bulk and {\it surface diffusion} which provides a mass transport and dissipative mechanism for crack propagation, are well established. 
The model is essentially parameter free, leading to the prediction that two dimensionless quantities, the crack velocity - $v/v_R$ and crack tip radius - $r_0(v_R/D)^{1/3}$ are universal functions of the dimensionless driving force $\Delta$.
Strictly speaking, these functions still depend weakly on the Poisson ratio.
We note that these statements, together with the prediction of the tip instability above some critical velocity, are based on the general structure of our theory and do not involve the specific modeling of the surface stresses. 
The specific results given in the figures should differ from exact solutions only quantitatively.

It is important to realize that our model does not contradict classical theories \cite{Freund90}, but contains more information.
This allows to calculate both the crack velocity and the fracture energy, while the classical theories predict only a relation between these quantities (the integral energy balance).
In our model, this energy balance is not fulfilled by approaching the Rayleigh speed limit with increasing driving force, as the classical theories predict, but by an increase of the fracture energy by tip blunting, which eventually leads to the tip instability.

The scale of velocity is set by $v_{R}$ and is independent on the diffusion coefficient.
However, the length scale of the tip is set by $(D / v_{R})^{1/3}$.
As we already noted, strong dissipation brings the local crack tip temperature close to the melting temperature.
The diffusion coefficient $D_{s}$ is then about $10^{-7} {\rm m}^{2}{\rm s}^{-1}$ \cite{Neumann72} and independent of outside temperature.
Consequently, the lengthscale of the tip is of order of atomic units. 
We hope, that even for these small scales our general, qualitative predictions remain correct.

\begin{acknowledgments}
We thank S. Iordanskii, V. I. Marchenko, H. M\"ul\-ler-Krumb\-haar, H. Schober, D. Temkin and H. Trinkaus for useful discussions.
This work has been supported by the Deutsche Forschungsgemeinschaft (Grant SPP 1120).
\end{acknowledgments}


\begin{thebibliography}{99}
\bibitem{Unpub} Details of the derivation will be given elsewhere.
\end{thebibliography}
\end{document}